\documentstyle{article}
\begin{document}
\parindent=10pt
\def\be{\begin{equation}}
\def\ee{\end{equation}}
\renewcommand{\baselinestretch}{1.2}

\large

\begin{center}

{\LARGE \bf The myth about nonlinear differential equations}

C. Radhakrishnan,

Centre for Mathematical Sciences

Pangappara, Thiruvananthapuram,

695 581 India

\large

September 13, 2001

\vskip1cm

{\bf Abstract}

\end{center}

Taking the example of Koretweg--de Vries equation, it is shown that soliton
solutions need not always be the consequence of the trade-off between the
nonlinear terms and the dispersive term in the nonlinear differential
equation. Even the ordinary one dimensional linear partial differential
equation can produce a soliton.

\vskip1cm

Solitary waves and solitons are often described [1-8] as a consequence of
the trade-off between nonlinear and dispersive terms in the nonlinear
differential equations. This has given rise to the myth that solitary
waves and solitons can be obtained as solutions of nonlinear differential
equations only and not as solutions of linear differential equations. An
associated misunderstanding is that only nonlinear differential equations
are capable of describing nonlinear physical phenomena and nonlinear
differential equations are more powerful than linear differential
equations in describing physical phenomena. The observation that, in
nature, linear phenomena are often only approximations to nonlinear
phenomena probably gave birth to this belief. As a consequence, physicists
are constructing more and more nonlinear differential equations to
describe nonlinear physical phenomena. It is, of course true that
nonlinear phenomena are more general, than linear phenomena. But linear
differential equations are, in general neither approximations nor
particular cases of nonlinear differential equations. Also, it is to be
emphasised that solutions of both linear and nonlinear differential
equations are functions which depend nonlinearly on the independent
variable (the only exception being the straight line solution). We can
construct linear as well as nonlinear differential equations from the same
function. For example, from the function
\be
        y(x,t) = x^2  +  t^2  
\ee
we can construct the linear second order partial differential equation
(PDE)

\be
\frac {\partial^2y} {\partial x^2} = \frac {\partial^2y} {\partial t^2}
\ee

and the first order, second degree nonlinear differential equation
\be
\left(\frac{\partial y}{\partial x}\right)^2 + \left(\frac{\partial y}{\partial t}\right)^2
= 4 y 
\ee
Conversely,
 $$               y(x,t) = x^2 +  t^2 $$
is a solution of the linear differential equation (2) and the nonlinear
differential equation (3). This simple example illustrates that the
distinction between linear and nonlinear differential equations appears
only in the differential equation level and not in the solution level.
Looking at equations (1) - (3), it is obvious that nonlinearity is hidden
in the second order linear differential equation (3) and the higher the
order of the differential equation, the higher the nonlinearity it
contains.

Any physical phenomena is described within a model. Within this model, the
physical phenomenon, in general, can be described with the help of a
linear differential equation or a nonlinear differential equation or a
function. The information content in the linear differential equation with
its initial and boundary conditions or in the equivalent nonlinear
differential equation with its initial and boundary conditions or in the
solution function will be the same. Therefore, the claim that a particular
physical phenomenon can be described only by a nonlinear differential
equation, and not by any linear differential equation is not tenable,
provided a linear differential equation with the same solution as that of
the nonlinear differential equation exists. The same philosophy, as above,
is at work, in the efforts of pure mathematicians to find linear operators
equivalent to nonlinear operators. Incidentally, linearization is the
oldest and most popular method of solving nonlinear differential
equations. Linearization is essentially the setting up of a linear
differential equation with its solutions the same as some of the solutions
of the nonlinear differential equation.

For the purpose of this letter, a linear differential equation and a
nonlinear differential equation having the same solutions are called
equivalent differential equations. Throughout this letter, to save space
and time, we use the Koretweg--de Vries (KdV) equation and its solutions in
the form given by Drazin and Johnson [1].

The KdV equation
\be
u_t - 6 u u_x + u_{xxx} = 0 
\ee
has the soliton solution
\be
u(x,t) = - { 1 \over 2}  csech^2 \left\{ {1 \over 2} c^{1 \over 2}
(x - ct - x_0) \right\}
\ee
It can be easily verified that (5) is also a solution of the first order
one dimensional wave equation
\be
u_x = - {1 \over c} u_t
\ee
Therefore, the nonlinear differential equation (4) or the linear
differential equation (6) or the solution (5) equally well represent the
soliton. Hence, the argument that the solitons are produced only as a
result of the trade-off between the nonlinear terms and the dispersive
term of the nonlinear differential equation is not valid at least, for the
soliton represented by the function (5). From the point of view of the
operator formalism, we can say that the nonlinear KdV operator,
$$ \frac {\partial} {\partial t} - 6 u \frac {\partial^3} {\partial x} 
+\frac {\partial} {\partial x^3} $$
and the linear operator
$$\frac {\partial} {\partial x} + {1 \over c}  \frac {\partial} {\partial t} $$
are identical so far as the function (5) is concerned. Equation (5) is
also a solution of the {\it n}th order wave equation
\be
\frac {\partial^n u} {\partial x^n} = (-1)^n \frac {1} {c^n}
\frac {\partial^n u} {\partial t^n}
\ee
for any value of n. Therefore, there are infinite number of linear
operators
$$ \frac {\partial} {\partial x} = - {1 \over c} \frac {\partial} {\partial t}
~~ , ~~~~~~ \frac {\partial^2} {\partial x^2} = \frac {- 1} {c^2} \frac {\partial^2} {\partial t^2}
 ~~ , ~~  \cdots  $$
equivalent to the nonlinear KdV operator for the soliton solution (5). 

The
singular soliton solution [1],
\be
u(x,t) = 2 k^2 cosech^2 \left\{ k(x - 4 k^2 t ) \right\}
\ee
%and the N-soliton solution [1]
%\be
%u(x,t) \sim -2 \Sigma n^2 sech^2 \left \{n (x - 4 n^2 t) \mp x_n \right\}
% as t - \pm \inf
%\ee
of the KdV equation (4) is also a solution of the nth order linear
differential equation (7) for any value of n.

Other nonlinear differential equations with soliton solutions have also
equivalent differential equations. Since one example is enough to
establish our point we do not cite further examples here.

As we have already mentioned, there may be soliton solutions for which a
linear PDE may not exist. As an interesting example for the rational
solution [1]
\be 
u(x,t) = \frac { 6 x (x^3 - 24 t) } { (x^3 + 12 t )^2}
\ee
of the KdV equation
\be
u_x = {1 \over \kappa} u_t ~~~~ {\rm where}~~ \kappa {\rm~~ is~~ a~~ constant}
\ee
is not an equivalent linear equation.

%\eject

It is interesting to point out in this connection that Davidov[9]  has
made the following observation about soiltons from linear differential 
equations:

``Solitonic-type solutions give a second stable branch of solutions
of the Schrodinger equation. In a certain sense, these solutions
are isolated from the boundary c onditions due to their localisation
on a rather small region of the chain. 

 A soliton is described by a wave whose profile $\phi(\xi)$ is unchanged
 under propagation. Such waves refer to the stationaty ones,
 for which the following relation holds.
 $$ \frac {\partial \phi } { \partial t} = - V \frac {\partial \phi } 
 {\partial x} $$
 where V is the velocity of propagation.''

 The  significance of this finding is that solitons and many other
 complex entities associated with non-linear physical phenomena
 can be studied, predicted, and described in terms of linear differential
 equations. The additional advantage of using linear differetial equations
 is that several techniques and tools have been developed in the
 past 300 years which can be effectively used to solve complex 
 non-linear phenomena.

\vskip1cm

{\bf Acknowledgements}

\vskip1cm

I have been benefitted by the criticism offered by G. Mohanachandran. I
wish to thank E. Krishnan for the preparation of the manuscript and
helpful discussions. The last touches to this letter were given during a 
short visit to TIFR, Mumbai. 

\vskip1cm

{\bf References}

\parindent0pt

\vskip1cm

[1] Drazin P G and Johnson R S 1990 Solitons : an Introduction
(Cambridge:Cambridge University Press)

[2] Tabor M 1989 Chaos and Integrability in Nonlinear Dynamics (New York:
John Wiley and Sons)

[3] Ashok Das 1989 Integrable Models (Singapore: World Scientific)

[4] Whitham G B 1974 Linear and Nonlinear Waves (New York: Wiley)

[5] Agrawal G P 1989 Nonlinear Fiber Optics (San Diego: Academic)

[6] Billingham J and King A C 2000 Wave Motion (Cambridge: Cambridge
University Press)

[7] Gapanov-Grekhov A V and Rabinovich M I 1992 Nonlinearities in Action
(Berlin: Springer)

[8] David Logan J 1997 Applied Mathematics (New York: John Wiley and Sons)
P305.

[9] Davidov A S 1991 Solitons in Molecular Systems, 2nd ed (Dordrecht,
Kluwer Academic Publishers) P31 -- 32

\end{document}